\documentclass[journal]{IEEEtran}
\usepackage{cite,url}

\ifCLASSINFOpdf
   \usepackage[pdftex]{graphicx}
\else
\fi
\usepackage{algorithm} 
\usepackage{algpseudocode} 
\usepackage{amsmath}
\usepackage{graphics} 
\usepackage{epsfig} 
\usepackage{mathptmx} 
\usepackage{amssymb}  
\usepackage[english]{babel}
\usepackage[utf8]{inputenc}
\usepackage{amsmath, amsfonts}
\usepackage{graphicx}
\usepackage{soul}
\usepackage{color}
\usepackage[dvipsnames]{xcolor}
\usepackage{breqn}
\newcommand{\comment}[1]{}

\renewcommand{\bar}{\overline}
\usepackage{caption}
\usepackage{subcaption}
\usepackage{titlesec}
\usepackage{soul}
\usepackage{tikz}

\newcommand\copyrighttext{%
  \footnotesize This work has been submitted to the IEEE for possible publication. Copyright may be transferred without notice, after which this version may no longer be accessible.}
\newcommand\copyrightnotice{%
\begin{tikzpicture}[remember picture,overlay]
\node[anchor=south,yshift=10pt] at (current page.south) {\fbox{\parbox{\dimexpr\textwidth-\fboxsep-\fboxrule\relax}{\copyrighttext}}};
\end{tikzpicture}%
}

\begin{document}

\title{MPC-based Realtime Power~System~Control with DNN-based Prediction/Sensitivity-Estimation}

\author{Ramij Raja Hossain,~\IEEEmembership{Student Member,~IEEE,}
        Ratnesh Kumar,~\IEEEmembership{Fellow,~IEEE}
\thanks{The work was supported in part by the National Science Foundation under
the grants, CCF-1331390, ECCS-1509420, PFI-1602089, and CSSI-2004766.
\newline \indent R. R. Hossain and R. Kumar are with the Department of Electrical and
Computer Engineering, Iowa State University, Ames, IA 50010, USA (e-mail:
rhossain@iastate.edu, rkumar@iastate.edu).}
\vspace*{-.3in}}

\maketitle
\copyrightnotice
\begin{abstract}
This paper presents a model predictive control (MPC)-based online real-time adaptive control scheme for emergency voltage control in power systems. Despite tremendous success in various applications, {\em real-time} implementation of MPC for control in power systems has not been successful due to its online computational burden for large-sized systems that takes more time than available between the two control decisions. This long-standing problem is addressed here by developing a novel MPC-based adaptive control framework which (i) adapts the nominal offline computed control, by successive control corrections, at each control decision point using the latest measurements, (ii) utilizes data-driven approach for prediction of voltage trajectory and its sensitivity with respect to control using trained deep neural networks (DNNs). 
In addition, a realistic coordination scheme among control inputs of static var compensators (SVC), load-shedding (LS), and load tap-changers (LTC) is presented with a goal of maintaining bus voltages within a predefined permissible range, where the delayed effect of LTC action is also incorporated in a novel way. The performance of the proposed scheme is validated for IEEE 9-bus as well as 39-bus systems, with $\pm 20\%$ variations in nominal loading conditions. We also show that the proposed new scheme speeds up the online computation by a factor of 20 bringing it down to under one-tenth the control interval, making the MPC-based power system control practically feasible. 

\end{abstract}



\section{Introduction}
\IEEEPARstart{R}{eal-time} control in power systems is of vital importance for resilience and security of the bulk power system with the increasing integration of renewable energy sources and dynamic loads. Although most of the power utilities are equipped with fast, robust and reliable protective relaying scheme, severe disturbances in power systems such as system faults, loss of generation, or circuit contingencies can cause \emph {large-disturbance voltage instability} resulting in a significant decline in bus voltages even after several seconds of fault clearance \cite{1318675}. 
This necessitates incorporation of Special Protection Systems (SPS) or Remedial Action Schemes (RAS) 
to exercise adaptive control actions under an event of potential instability. Standard practices generally include empirical rule-based approach using predetermined control setting \cite{1270926} to maintain voltage trajectories within specified limits, but these approaches fail to adapt changing operating conditions and are not suitable in modern power systems with uncertain load and generation profiles. To this end, MPC (Model-Predictive Control) is a promising alternative for traditional SPS-based control in power system. The existing rich {\em theoretical} study on MPC in power system application \cite{larsson2003coordinated,wen2004optimal, zima2005model,hiskens_mpc,cutsem_review,jin2009model} points to the possibility for this kind of control scheme, but its real-time implementation remains to be demonstrated. 

The computational complexity of online optimization is the major hindrance of MPC for real-time application in power system control, which requires each control action computation time to be significantly less than the control interval. Recent advances in data-driven methods have proposed deep reinforcement learning (DRL) based control schemes, that attempt to address the issues of computational burden. DRL-based schemes, generally, optimize an abstract reward signal to train the control agent offline without accessing accurate model information \cite{li,huang}. There are certain issues associated with these DRL schemes: (i)  it is often hard to train an agent in model-free manner for power system dynamics considering its complexity, (ii) also the design of reward function is non-trivial, and its design is instrumental for successful training of a DRL-agent. Moreover, the system operators have always been concerned about safety and reliability of the control schemes, and only those control techniques that are fully understood, matured, and are guaranteed to work are utilized \cite{cutsem_review}. In contrast to DRL, MPC fully depends on the model information and iteratively computes a sequence of control variable adjustments at each control instant optimizing a predicted future behavior cost of the underlying system. It implements the first action of the computed sequence at the current control instant, and repeats the process with new measurement at each next control instant \cite{garcia1989model,jin2009model}. 
To be able to utilize the advantages of well-accepted MPC in power systems, it is imperative to significantly reduce the online computational time (to a small fraction of the time between any to control decision instants). 

To understand the time requirement for each MPC iteration, one can note that it includes: measurement of current state, prediction of future trajectory, and solving of optimization to compute the required control adjustment. Due to the presence of non-linearity and complexity of power systems, the most time consuming step is the prediction of change in future trajectory as a function of change in controls. One of the well established approach is trajectory sensitivity \cite{hiskens2000trajectory} which involves linearizing the system around a nominal trajectory rather than, as is more customary, around an equilibrium point, making the former more accurate. The traditional computation of trajectory sensitivity needs the full-blown time domain simulation of system dynamics, which in itself is time consuming for power systems, increasing exponentially with the size of the power network, and making the realization of MPC impractical in real-world setting. This paper addresses the issues of real-time implementation of MPC with the following specific contributions:

\begin{enumerate}
     \item An approach to real-time MPC that reduces the online computation time to a small fraction of the control interval. Firstly, an optimal control sequence is generated solving a standard MPC formulation utilizing the nominal model of the underlying power system, and stored. It is understood that this generated sequence is optimal only for the nominal model, and so in our approach it is corrected successively online against the real-time system, based on its measurements (as in neighboring optimal control or perturbation control \cite{lewis2012optimal}).
     Furthermore, the real-time update is achieved by solving a proposed predictive optimization, where the predicted trajectory and corresponding sensitivities are determined using trained DNNs. The incorporation of DNN and online control correction greatly speeds up the overall online computation, showing a gain by a factor of 20 in online computation time, and bringing the computation to a fraction of the control decision time, making the MPC-based scheme practical for real-world application.
    \item We develop and present an MPC-based voltage control scheme with realistic coordination among SVC, LS and LTC, where the slow acting LTCs are formulated to have delayed control effect (delayed by two control periods) to make those better replicate their real-world behavior. Also, an LTC decision action is not issued unless it is determined that the corresponding condition remains to be satisfied for the two future control decision instants.
    
    \item A comprehensive and accurate framework for the computation of trajectory sensitivities with respect to control inputs of SVC, LTC, and LS is developed and implemented in PSAT/MATLAB platform. 

    \item We demonstrate the performance of our approach with IEEE 9-bus and 39-bus power system models. The performance and robustness of the framework is validated under load variations of $\pm 20 \%$ to the nominal model. An improvement in computation speed by a factor of 20 is demonstrated for both of the systems, bringing down the MPC computation time to one-tenth of the control interval, and making the MPC-based control real-time as well as practical for real-world power system application for a first time. 
\end{enumerate}

\subsection{Related Works} \label{1A}
The idea of predictive control has been studied for decades in power systems. In \cite{larsson2003coordinated}, a coordinated voltage stabilization scheme using load shedding, capacitor switching and tap changers based on predictive control and tree search is presented. \cite{wen2004optimal} presents optimal voltage control using MPC, where voltage trajectories for selected control actions are predicted using a single-stage Euler state predictor (SESP). This work utilizes pseudo gradient evolutionary programming (PGEP) technique for solving optimization. In both cases, the optimization is combinatorial, hence is not scalable in large control-space for real-time application. 
In \cite{zima2005model}, a receding horizon technique is employed with the control actions load shedding and regulator voltage reference. An MPC based voltage control strategy is proposed in \cite{hiskens_mpc}, where the objective is to find optimized non-disruptive load control to prevent voltage collapse. \cite{cutsem_review} provides a reflection on early MPC-based approaches in power system, and also proposed quasi steady state (QSS) model and sensitivity based MPC-scheme combining a static and dynamic optimization. 

Among the prior works from our group, \cite{jin2009model} implemented a trajectory sensitivity based MPC-scheme utilizing only shunt capacitors. In this work, trajectory deviation and the cost of controls are simultaneously minimized. \cite{jin2009coordinated} provided an extension of the previous work with an MPC-based coordinated voltage control considering shunt capacitors, load-shedding and LTC operating, all at same time-scales. \cite{p1} proposed an offline planning approach of finding the amount and location of optimal reactive compensations considering a set of contingencies. An improvement in computation time of trajectory sensitivity is found in \cite{guanji,guanji2,Guanji3}; these works adopted a so called ``very dishonest newton (VDHN)" method to update Jacobians required for sensitivity computation. The use of VDHN introduced approximation and also their implementation remains unscalable beyond a short prediction horizon, and hence not suitable for generic real-world application with longer prediction horizon. Among recent works, \cite{zhang2018power}, proposed an adaptive horizon based MPC scheme in voltage control utilizing the idea of trajectory sensitivity. The main focus of this work is to find the optimal setting of horizon parameter using an predefined evaluation index and measurements. \cite{haomin1,haomin2}, proposed adaptive coordinated control using an offline-online approach. The online learning adopted a genetic algorithm (GA)-based exploration, which is time intensive for emergency control in real-world implementation. The application of MPC-based schemes are also becoming popular in active distribution network in presence of distributed generation and storage systems \cite{valverde2013model,aristidou2015contribution,guo2018mpc}. These works have also incorporated the sensitivity based implementation to formulate the optimization problem. Besides, there are also growing interest in distributed-MPC based approaches in power systems \cite{venkat2008distributed,moradzadeh2012voltage,rrh,9240947}. 

The extensive research for decades demonstrates that MPC-based methods can provide potentially good solutions from theoretical perspective. They become practically meaningful only if their computational bottleneck is addressed efficiently. Expicit-MPC (eMPC), a computationally viable alternative to traditional MPC, pre-computes the solution offline using multi-parametric programming (mpP) \cite{kvasnica2009real}.
But, eMPC suffers from the problem of 'curse of dimensionality' and can only be practical for small scale simple dynamical systems \cite{alessio2009survey}. Our proposed data-driven and online successive correction of offline computed nominal control policy offers a practical solution for real-time MPC-based centralized control for wide area power network for a first time.

\section{{Generic MPC Formulation for Power System}}\label{genericmpc}
The dynamics of power system is modeled using Differential Algebraic Equations (DAEs) of the form:
 \begin{equation}\label{sysdae}
 \dot{x}=f(x,y,u);\;\; 0=g(x,y,u),\;\;(x(t_0),y(t_0),u(t_0)) \mbox{ as given},
       \end{equation}
\noindent where we have, $x:=$ state variables, $y:=$ algebraic variables, and $u:=$ control inputs. In general the effect of any small disturbances in power system can be studied by linearizing (\ref{sysdae}) with respect to the current equilibrium point. But, following any large disturbances (line fault, generator outage etc.), which can cause a shift in the existing equilibrium, it is necessary to consider the complete DAE model for studying post disturbance system behavior (and not just its linearization around the pre-fault equilibrium). Accordingly, designing optimal control to improve system performance following large disturbances, requires the inclusion of (\ref{sysdae}) as a constraint in the optimization process. Thus one can define a general non-linear optimal control problem for power systems as follows: 
\begin{subequations}
\begin{equation}\label{genmpc1}
\min_{{u({\cdot})}}\int_{t_k}^{t_{k+T}} {l({x}(\tau), {y}(\tau), {u}(\tau))d\tau}
 \end{equation}
subject to:
\begin{equation}\label{genmpc2}
 \dot{{x}}=f({x},{y},{u}),\;\;\;
  0=g({x},{y},{u}),
  \end{equation}
     \begin{equation}\label{genmpc3}
x(t_k) = x_k\;,\;y(t_k) = y_k,
       \end{equation}
\begin{equation}\label{genmpc4}
x \in X,\;\; y \in Y,\;\; u \in U.
\end{equation}
\end{subequations}
The solution of the optimal control problem defined in (\ref{genmpc1})-(\ref{genmpc4}) can be obtained using Dynamic Programming (DP) which deals with the exact information about the future of the optimal trajectories. This in general makes DP a very hard problem to solve. In power systems, where dimensions of $f$ and $g$ are large, the problem becomes impractical. 

To this end, MPC offers a practical way of solving (\ref{genmpc1})-(\ref{genmpc4}) by approximating the system model (\ref{genmpc2}). However, owing to complex dynamics of power systems, it is difficult to approximate (\ref{genmpc2}) by its step-response or impulse response model, as is extensively used in Dynamic Matrix Control (DMC) and Model Algorithmic Control (MAC), two basic formulation of MPC \cite{garcia1989model}. This leads to trajectory sensitivity based approximations that are commonly used in power systems, providing a reasonable approximation of complex system model given by (\ref{sysdae}). The idea behind  trajectory sensitivity is time-dependent linear approximation to quantify the impact of the control variations on the nominal trajectories of system variables $x(t)$ and $y(t)$. Accordingly, the sensitivities of the trajectories to the control changes are expressed as: 
\[S^x(t):=x_u(t)=\frac{\partial x(t)}{\partial u(t)} \mbox{ and } S^y(t):=y_u(t)=\frac{\partial y(t)}{\partial u(t)},\] 
and the dynamics of $x_u(t)$ and $y_u(t)$ is obtained by differentiating  (\ref{sysdae}) with respect to control input $u(t)$, providing:
\begin{subequations}
  \begin{align}\label{systs1}
 \dot{x_u}(t)=f_x(t){x_u(t)}+f_y(t){y_u(t)}+f_u(t),
 \\
 0=g_x(t){x_u(t)}+g_y(t){y_u(t)}+g_u(t).\label{systs2}
 \end{align}
\end{subequations}
The knowledge of the trajectory sensitivities $[x_u(t),y_u(t)]$ is then used to estimate the predicated trajectories $[\hat x(t),\hat y(t)]$ when a small control correction $u(t)$ is introduced to the nominal system, that has the nominal trajectory $[\bar x(t),\bar y(t)]$:
\begin{equation}\label{tspred}
\hat x(t)\approx\bar x(t)+S^x(t) u(t),\quad \hat y(t)\approx\bar y(t)+S^y(t) u(t).
\end{equation}
\noindent It is important to note that in the process of time domain simulation of (\ref{sysdae}), that provides $[\bar x(t),\bar y(t)]$, trajectory sensitivities $[x_u(t),y_u(t)]$ can also be obtained by solving (\ref{systs1})-(\ref{systs2}). Later, in Section \ref{sectscomp}, this computation procedure will be discussed in detail.

To further formulate the problem in a practical setting, where computations occur at discrete points in time, separated by sampling duration, denoted  $T_{s}$, one replaces the integration/differentiation by numerical version. Also, the manipulated input $u(t)$ is held constant over control interval, denoted here as $T_c$, with $T_c =M T_{s}$ where $M\geq 1$ is the number of sample instances between any two control instants (for example if sample period is 0.1~s and control decision is taken every 3~s, then $M=30$). We provide below a discretized MPC formulation where at any discrete control instant $k\geq 0$, MPC can evaluate $N_k\geq 0$ control decisions ($N_k$ decrease by 1 each time $k$ increases by 1, i.e., the case of receding control horizon), while optimizing the system behavior for a prediction horizon of $N$, where $N>N_k,\forall k$. At a control instant $k\geq 0$, the controller computation solves the following optimization with respect to the sequence of length $N_k$ of control variables, $u_{k,\rm seq}:=u_{k}, u_{k+1}, \cdots u_{k+N_k-1}$. 

Since sampling occurs at a faster time scale than the control decision occurs ($M=\frac{T_c}{T_s}$ times faster), for any variable $r$ and control time index $j\geq 0$, we introduce a ``range" notation $r_{j:j+1}$ to represent the set of sample values the variable $r$ takes between the control decision instants $j$ and $j+1$, namely, 
\[r_{j:j+1}:=\left[
\begin{array}{l} r_j\equiv r(jT_c)\\ r(jT_c+T_s)\\  \vdots \\ r(jT_c+(M-1)T_s)=r((j+1)T_c-T_s)
\end{array}\right] .\]
Using this notation, we have the following discrete-time version of the optimization at each control instant $k\geq 0$:
\begin{subequations}
\begin{equation}\label{genmpc11}
\min_{{u_{k}, \cdots, u_{k+N_k-1}}}
\;\;\;\sum_{i=0}^{N-1} {l(\hat x_{k+i:k+i+1}, \hat y_{k+i:k+i+1})}
 \end{equation}
subject to:
  \begin{align}\label{genmpc12}
\hat x_{k+i:k+i+1}=\bar  x_{k+i:k+i+1} + S^x_{{k+i:k+i+1}}\!\!\sum_{j=0}^{\min(i,N_{k-1})}\!\!{ u_{k+j}},\\
\quad \hat y_{k+i:k+i+1}=\bar  y_{k+i:k+i+1} + S^y_{{k+i-1:k+i}}\!\!\sum_{j=0}^{\min(i,N_{k-1})}\!\!{ u_{k+j}},
  \end{align}
\begin{equation}\label{genmpc14}
\hat x_{k+i:k+i+1} \in X,\;\; \hat y_{k+i:k+i+1} \in Y,\;\;\;\forall i\in[0,N-1],
\end{equation} 
\begin{equation}\label{genmpc13}
u_{k+j} \in {U},\;\;\;\forall j\in[0,N_k-1],
\end{equation}
\comment{\begin{equation}\label{genmpc14}
\hat x(k+i|k) \in X,\;\; \hat y(k+i|k) \in Y,\;\;\;\forall i\in[1,N]
\end{equation}}
\end{subequations}
\comment{As the control inputs are held constant in between two consecutive control instant, for notational simplicity, we can use following.
\begin{align*}
\hat x(k+i|k):=\hat x(k+i-1:k+i|k)\\
\hat y(k+i|k):=\hat y(k+i-1:k+i|k)
\end{align*}}
where recall that the variables $\bar x,\bar y$ represent the nominal state and algebraic variables under the nominal control, whereas the variables $\hat x,\hat y$ represent the state and algebraic variables under the indicated controls added sequentially, so at instant $k+i,i\in[0,N_k-1]$, the cumulative added control is $\sum_{j=0}^{\min(i,N_{k-1})} u_{k+j}$. 
Also as it is customary with MPC, at each control instant $k\geq 0$, the very first move $u_k$ of the computed control sequence is implemented, and the controller continues to repeat the same above type of optimization with the updated measurements of system variables at each next control instant. This iterative procedure constitutes an implicit feedback (where the most recent measurements are used to adapt the control decisions) and is of utmost importance in reducing the effect of modeling and measurement imperfections. 

\section{Theoretical Framework for MPC-Based Coordinated Voltage Control}\label{COORMPC}
\begin{figure}[htbp]
  \centering
    \includegraphics[width=0.45\textwidth]{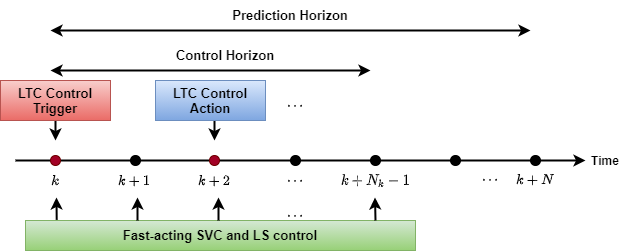}
  \caption{Illustrating principle of control coordination.}
  \label{f1}
 \end{figure}

In this section, we provide the theoretical formulation of MPC-based coordinated voltage stabilization problem of transmission network following a large disturbance, e.g., a line fault. It is often observed that even after clearance of fault, voltage trajectories may diverge from stable values due to the effect of load transients and other inherent dynamics of the power system. Hence, to stabilize the voltage trajectories, coordinated management of various controllable devices (Static VAR Compensators (SVC),  Under Voltage Load Shedding (UVLS), Load tap changers (LTC)) are required. However, as noted in introduction, the problem of an appropriate coordination among the controllers of {\em different time-scales} in the MPC setting hasn't been provided: For example, LTC, which are commonly used with the transformers at the boundary of transmission and distribution network, are slow-time scale devices compared to the fast-acting SVC or UVLS relays. The effect of the disturbance in a transmission network can propagate to the distribution side, that may prompt a tap setting change in LTC, commanded by its local automatic voltage control (AVC) system. The mechanical time delay $T_{\text{mech}}$ (typically $\sim 5$ s) is required for the LTC to move the taps by one position. Since $\lceil T_{\text{mech}}/T_c\rceil= 2$, the effect of the LTC comes after a delay of 2 control instants (as also illustrated in Figure~\ref{f1}) and this delay must be accounted. We present a practical MPC framework for coordinated voltage control involving shunt compensation and load-shedding while also considering the impact of the delayed actions of LTC. 

Without loss of generality, let us consider the distribution systems are modeled as a lump loads connected with the transmission side through transformers equipped with LTCs. In the proposed setting, the controls of SVC and LS on the transmission side are achieved by the voltage trajectory sensitivity based MPC, whereas the LTC control is achieved through local AVCs depending on the secondary side or low-voltage (LV) side bus voltages of respective transformers. We employ the decision logic of LTC control \cite{van1998voltage}, with an improvement to account for the $T_{\text{mech}}$ delay in LTC action: LTC performs a tap-change by a unit step at a control decision instant $k\geq 0$ if the predicted voltage of the low-voltage side controlled bus, $\hat{V}^{\text{lv}}$, ends up violating a predefined dead-band threshold, $V^{\text{db}}$, wrt a reference voltage $V^{\text{r}}$ over the period starting form the current control instant $kT_c$ and extending $T_{\text{mech}}$ units later. Given that in practical settiings, $\lceil T_{\text{mech}}/T_c\rceil= 2$, AVC needs to check the dead-band constraint for the predicted voltage over the control instants from $k$ to $k+2$ (i.e., for the variable $\hat{V}^{\text{lv}}_{k:k+2}$ in our ``range" notation) to adjust the tap setting:
\begin{equation}\label{eqnltc}
\Delta{\mathbf{N}_k} = 
\begin{cases}
   +1, & \text{if}\;\;\hat{V}^{\text{lv}}_{k:k+2}\leq V^{\text{r}}-\frac{V^{\text{db}}}{2}\\
   -1, & \text{if}\;\;\hat{V}^{\text{lv}}_{k:k+2}\geq V^{\text{r}}+\frac{V^{\text{db}}}{2}\\
   \;\;\;0, & \text{otherwise}.
\end{cases}
\end{equation}
\noindent The following observations are made about the LTC control:
\begin{itemize}
    \item As discussed above and illustrated in  Figure \ref{f1}, if a LTC change decision is taken at control instant $k$ according to (\ref{eqnltc}), it gets implemented two control instants later, at instant $k+2$. 
    \item Accordingly, the impact of the LTC decision at instant $k$ is captured in MPC optimization at instant $k+2$ onwards. 
    \item The above also implies that if a LTC change decision is taken at instant $k$, then the earliest next instant an LTC change decision can be taken is at instant $k+2$ (i.e., no LTC decision at instant $k+1$). We incorporate this feature in our formulation (see (\ref{genmpcV2})).
\end{itemize}
Next, we formalize the MPC optimization problem at a control instant $k\geq 0$ for a power system voltage control comprising $\mathbf{N}_b$ no. of buses. The formulation is derived from the general one given in (\ref{genmpc11})-(\ref{genmpc14}) considering only the trajectory of vector of voltage variable, $V := [V^1,\cdots,V^{\mathbf{N}_b}]^T$, where in (\ref{sysdae}) $V \in y$ is among the algebraic variables. For notational simplicity, we define the control-sequences, $u^ {\text{SVC}}_{{k,\text{seq}}}:=u^{\text{SVC}}_{k},\cdots,u^{ \text{SVC}}_{k+N_k-1}$ and $u^ {\text{LS}}_{{k,\text{seq}}}:=u^{\text{LS}}_{k},\cdots,u^{\text{LS}}_{k+N_k-1}$. 
\begin{subequations}
\begin{multline}\label{genmpcV1}
\min_{{u^{\text{SVC}}_{{k,\text{seq}}}\;\;,\;\;u^ {\text{LS}}_{{k,\text{seq}}}}}\;\;\;\sum_{i=0}^{N-1} (\hat V_{k+i:k+i+1} - V_{\text{ref}})^T \mathbf{R} (\hat V_{k+i:k+i+1} - V_{\text{ref}}) \\ \;+\; W^T_{\text{SVC}} {u}^{\text{SVC}}_{k,\text{seq}} \;+\;  W^T_{\text{LS}} {u}^{\text{LS}}_{k,\text{seq}}
\end{multline}
subject to,
\begin{eqnarray*}
\forall i\!\in\![0,N-1]\!:\hat V_{k+i:k+i+1}\!=\!\bar  V_{k+i:k+i+1}\!+\! S^{\text{SVC}}_{{k+i:k+i+1}}\times\!\!\!\!\!\!\!\!\!\!\!\!\sum_{j=0}^{\min(i,N_k-1)}\!\!\!\!\!\!{u^ {\text{SVC}}_{k+j}} \\ +\;\; S^{\text{LS}}_{{k+i:k+i+1}}\times\!\!\!\!\!\!\!\!\!\!\!\!\sum_{j=1}^{\min(i,N_k-1)}\!\!\!\!\!\!{u^{ \text{LS}}_{k+j}}\;\; +\;\; \mathbf{I} \times S^{\text{LTC}}_{{k+i:k+i+1}}\times{u^{\text{LTC}}_k}, 
\text{ where\;\;}
\end{eqnarray*}
\begin{equation}\label{genmpcV2}
\mathbf{I} =
\begin{cases}
      0, & \text{if\;\;$k$ is not a LTC decision point}\\
      0, & \text{if\;\;$k$ is a LTC decision point and } i<2\\
  1, & \text{otherwise}.
\end{cases}
  \end{equation} 
\comment{\begin{equation*}
\mathbf{I} = 
\begin{cases}
   0, & \text{if} $k$\;\;i\leq 2\\
   1, & \text{otherwise}.
\end{cases} \text{OR,\;\;\;\;} \mathbf{I} = 
\begin{cases}
   0, & \text{if}\;\;i\leq 1\\
   1, & \text{otherwise}.
\end{cases}
\end{equation*}}
 \begin{equation}\label{genmpcV3}
u^{\text{SVC}}_{\text{min}} \leq u^{{\text{SVC}}_{k+j}} \leq u^{\text{SVC}}_{\text{max}},\;\;\;\forall j\in[0,N_k-1]
\end{equation}
\begin{equation}\label{genmpcV4}
u^{\text{LS}}_{\text{min}} \leq u^{{\text{LS}}_{k+j}} \leq u^{\text{LS}}_{\text{max}},\;\;\;\forall j\in[0,N_k-1]
\end{equation}
\begin{equation}\label{genmpcV5}
V_{\text{min}} \leq \hat V_{k+N} \leq V_{\text{max}} 
\end{equation}
\end{subequations}
Here:
\begin{itemize}
    \item $V_{\text {ref}}:=$ reference voltage, $\mathbf{R}:=$ weight matrix for bus voltage deviation, $W_{\text{SVC}}$ and $W_{\text{LTC}}$ := weight vectors for SVC and LS control inputs, respectively.
    \item  $S^{\text{SVC}},S^{\text{LS}}$, and $S^{\text{LTC}}:=$ Voltage trajectory sensitivity matrices wrt. SVC, LS and LTC control input, respectively,
    \item $u^{{\text{LTC}}}_k \;\;:=\;\;\Delta{\mathbf{N}}_k\times \Delta V_{\text{tap}}$, where $\Delta{\mathbf{N}}_k:=$ LTC control decision made at $k$, and $\Delta V_{\text{tap}}:=$ p.u. voltage change per tap operation.
    \item $[u^{\text{SVC}}_{\text{min}},u^{\text{SVC}}_{\text{max}}]\;,\; [u^{\text{LS}}_{\text{min}},u^{\text{LS}}_{\text{max}}]\;,\; [V_{\text{min}},V_{\text{max}}]:=$ Lower and upper bounds for changes in SVC, LS control inputs, and the voltage values at the end of prediction horizon, respectively.
\end{itemize}

\comment{$V_{\text {ref}}:=$ reference voltage, $S^{\text{SVC}},S^{\text{LS}}$, and $S^{\text{LTC}}$ := voltage trajectory sensitivity matrices wrt. SVC, LS and LTC control input, respectively. $\Delta{\text{LTC}}:=$ LTC control decision made at $k$ }

The optimization problem defined in (\ref{genmpcV1})-(\ref{genmpcV5}) has a quadratic objective/cost function with linear inequality constraints, making it a quadratic programming (QP), which can be efficiently solved for the nominal system. A positive-definite matrix choice of $\mathbf{R} > 0$ makes the optimization problem convex, ensuring a global optimal solution. It is important to note that LTC action is taken based on the dead-band requirement of (\ref{eqnltc}), and the impact of this action on the voltage trajectory computation is incorporated within the MPC optimization. This approach (of LTC action) is supported by the work in \cite{guo2018mpc}, which analyzes MPC-based voltage regulation in distribution network with Distributed Generation (DG) and Energy Storage Systems (ESS). This LTC action approach also lends computational benefits since optimization over discrete control variable of LTC is not required, rather implemented per (\ref{eqnltc}).

\section{{Online Adaptive Control: Control corrections to Nominal Offline control}}\label{secpertcontrol}

In Section~\ref{genericmpc} and \ref{COORMPC}, we provided a theoretical framework for MPC-based voltage control in power systems. To translate the theory into real-time implementation, the following tasks need to be accomplished in runtime at each control instant $k\geq 0$:
\begin{enumerate}
    \item Real-time measurement of system variables
    \item Model-based prediction of future trajectory and its sensitivity computation using the measured values
    \item Solving optimization problem (\ref{genmpcV1})-(\ref{genmpcV5})
\end{enumerate}
For any practical power system, the step 2), namely, model based trajectory prediction and trajectory sensitivity computation is computationally expensive for a practical prediction horizon (20-50 s), and hence it is often infeasible to perform MPC optimization in the desired time-frame of control computation (needs to be much smaller than $T_c\sim\!3$ s). To this end, one possible direction is to explore artificial intelligence particularly the advances in deep learning to train for the trajectory prediction and sensitivity computation for the sake of reducing their online computational time-complexity. 

Here the main challenges lies in the size of training set required for offline training of the required DNNs for trajectory prediction and sensitivity computation. A simple analysis shows that in a MPC of $N_c$ control steps with $m$-dimensional controls, each having $q$ levels of quantization, the number of possible  
control combinations is $q^{m\times N_c}$, 
and if there are $p$ number of load/contingency scenarios, then total combinations to explore for training is $p\times q^{m\times N_c}$, which is prohibitive. 
To overcome this prohibitive size of training space, we introduce a novel approach involving successive control correction at each control instant, limiting the training space to the neighborhood of a nominal-case optimal trajectory for training the DNNs, which brings down the training space size to order $p$ (since the control choice is already fixed at the nominal values found in the offline MPC optimization). This reduction in training space helps make it practically feasible.

\begin{figure}[htbp]
  \centering
    \hspace*{-.1in}\includegraphics[width=0.50\textwidth]{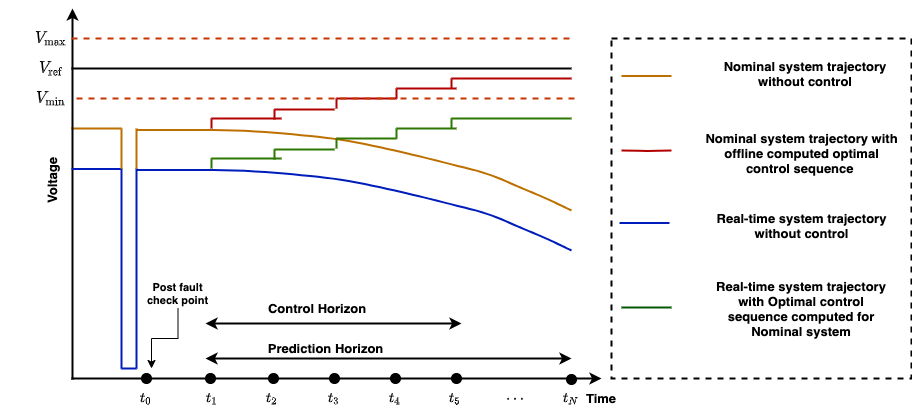}
  \caption{Illustrating online control correction computation.}
  \label{f2}
  \vspace*{-0.1in}
 \end{figure}

In the proposed approach, one views the real system trajectory to be a corrected version of the nominal system trajectory, 
that is obtained by an offline MPC-based optimization of the given nominal system model.
For example, the real-time operation load levels vary within $\pm 20\%$ of the nominal values of load. 
Figure \ref{f2} presents an illustration. The \textbf{yellow} curve and the \textbf{blue} curve are post fault trajectory of the nominal system versus the real system without any control. The offline computed optimal control sequence $u^*_{\text{nom,seq}} := u^*_{1,\text{nom}},\cdots,u^*_{N_c,\text{nom}}$ (with $N_c$ control steps) can stabilize voltages of the nominal system within the desired range (see \textbf{red} curve), but the same is not true for the real system (see \textbf{green} curve). Hence a correction in the nominal control is required.
Suppose $u^*_{\text{nom,seq}} := u^*_{1,\text{nom}},\cdots,u^*_{N_c,\text{nom}}$ is the offline computed MPC-based optimal control sequence for the nominal system following a particular contingency.  
Then the optimal control sequence for the same contingency for the actual real system $u^*_{\text{real,seq}}$, which differs from the nominal system model, can be written as a correction $\Delta u^*_{\rm seq}$ to $u^*_{\text{nom,seq}}$:
\begin{equation}\label{pertcont}
u^*_{\text{real,seq}}\; = \;u^*_{\text{nom,seq}} \;+ \;\Delta u^*_{\rm seq}\;. 
\end{equation}

In our approach, the sequence of control corrections $\Delta u^*_{\rm seq} := \Delta u^*_1,\cdots,\Delta u^*_{N_c}$ in (\ref{pertcont}) is estimated online utilizing the measurements of the real system, while meeting the computational time constraints imposed by the real-time operation of a power system. Next we outline our approach for the computation of the corrections in control in a iterative manner at the control instants, where to compute the desired control correction term at control instant $k\geq 0$, we propose the following steps:
\begin{itemize}
    \item[(a)] First, predict the voltage trajectory of the real system under the influence of $u^*_{k,\text{nom}}$ over the time range $kT_c$ to $(k+1)T_c-T_s$ (see description below).
    \item[(b)] Next, compute the sensitivities of the voltage trajectory obtained in (a) with respect to controls (see described below).
    \item[(c)] Using the information obtained in (a) and (b), solve a 1-step MPC optimization to compute $\Delta u_k^*$.
    \item[(d)] Repeat the above steps at the next control instant, until the final control instant is reached.
\end{itemize}

\begin{figure}[htbp]
  \centering
    \includegraphics[width=0.48\textwidth]{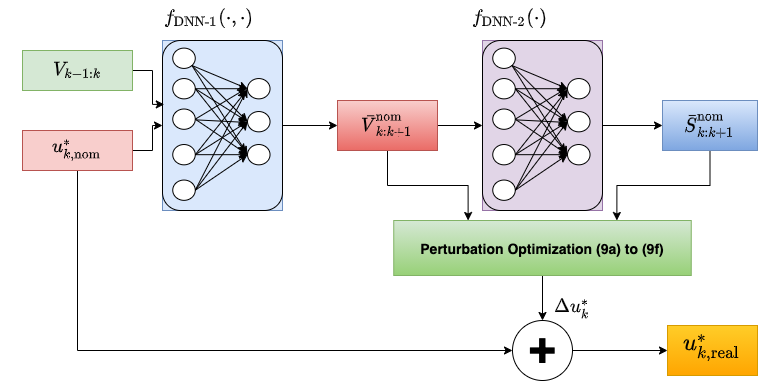}
  \caption{Iterative online computation of control correction}
  \label{f3}
 \end{figure}

To facilitate the trajectory prediction and sensitivity computation in steps (a) and (b) in real-time fashion, we introduce two trained DNNs (see Figure~\ref{f3}). This eliminates the necessity of time-consuming model based online time-domain simulation for trajectory prediction and its sensitivity estimation, making the online computation of control corrections viable for real-time MPC application. 
\begin{itemize}
\item {\bf The Prediction-DNN, $f_{\text{DNN-1}}(\cdot,\cdot)$:} At each control instant $k\geq 0$, it receives as inputs (i) the measured voltage trajectory of the real system $V_{k-1:k}$ between the last two control instants (from $(k-1)T_c$ to $kT_c-T_s$), and (ii) the nominal optimal control at $kT_c$, $u^*_{k,\text{nom}}$, while it outputs the predicted trajectory $\bar V^{\rm nom}_{k:k+1}$ over $kT_c$ to $(k+1)T_c-T_s$. The idea here is that the measured voltage trajectory $V_{k-1:k}$ over the period $(k-1)T_c$ to $kT_c-T_s$ serves as the proxy to the state at $kT_c$, which when combined with the current control information $u^*_{k,\text{nom}}$ at $kT_c$, allows the prediction of the future state proxy, namely, the voltage trajectory $\bar V^{\rm nom}_{k:k+1}$ over the period $kT_c$ to $(k+1)T_s-T_s$ under the nominal control.

\item {\bf The Sensitivity-DNN, $f_{\text{DNN-2}}(\cdot)$:} It receives the predicted trajectory, i.e., the output $\bar V^{\rm nom}_{k:k+1}$ of  $f_{\text{DNN-1}}(\cdot,\cdot)$ as input, and produces the sensitivity matrix $\bar S^{\rm nom}_{k:k+1}$ as its output. 
\end{itemize}
 
Next, we introduce the single step MPC optimization that we propose to solve at each control instant $k\geq 0$ towards the step (c) above. \begin{subequations}
\begin{equation}\label{pertmpcV1}
\min_{\Delta u_k}\;\;\;{(\hat V_{k:k+1} - V_{\text{ref}})^T \mathbf{R}_{\Delta} (\hat V_{k:k+1} - V_{\text{ref}})} + w_{\Delta}^T \Delta u_k 
\end{equation}
subject to,
  \begin{equation}\label{pertmpcV2}
\hat V_{k:k+1} = \bar V^{\text{nom}}_{k:k+1} + \bar S^{\text{nom}}_{k:k+1}\;\Delta u_k
  \end{equation}
 \begin{equation}\label{pertmpcV3}
\bar V^{\text{nom}}_{k:k+1} = f_{\text{DNN-1}}(V_{k-1:k}, u^*_{k,\text{nom}})
\end{equation}
 \begin{equation}\label{pertmpcV4}
\bar S^{\text{nom}}_{k:k+1} = f_{\text{DNN-2}}(\bar V^{\text{nom}}_{k:k+1})
\end{equation}
 \begin{equation}\label{pertmpcV5}
u_{\text{min}} \leq u^*_{k,\text{nom}} + \Delta u_k \leq u_{\text{max}}
\end{equation}
\begin{equation}\label{pertmpcV6}
V_{k,\text{min}} \leq \hat V_{k+1} \leq V_{k,\text{max}} 
\end{equation}
\end{subequations}
\noindent where 
$\mathbf{R}_{\Delta}$ and $w_{\Delta}:=$ appropriately truncated portions of $\mathbf{R}$ and $[W_{\rm SVC}^T\; W_{\rm LS}^T]$ respectively to account for the single step costs, $\bar V^{\text{nom}}_{k:k+1} :=$ predicted voltage trajectory over $kT_c$ to $(k+1)T_c-T_s$ under nominal optimal control using DNN-1, and $\bar S^{\text{nom}}_{k:k+1}:=$ predicted sensitivity over $kT_c$ to $(k+1)T_c-T_s$ under nominal optimal control using DNN-2. The output of the optimization problem (\ref{pertmpcV1})-(\ref{pertmpcV6}) gives $\Delta u^*_k$, which is then added to $u^*_{k,\text{nom}}$ to get $u^*_{k,\text{real}}$ as in (\ref{pertcont}), which is then implemented at $k$.

As per our previous discussion, the control input $u$ of our setting has 3 different component corresponding to SVC, LS and LTC. We remind the LTC decision and its implementation: At a control decision point $k\geq 0$, the local AVC controllers predict the LV side voltage trajectories of designated buses over the control instants $k$ to $k+2$ (from $kT_c$ to $(k+2)T_c-T_s$) and issue control decisions according to (\ref{eqnltc}). These decisions are implemented at $k+2$ to account for the delayed action of LTC. For a real-time prediction of future LV voltage, we introduce another DNN, an {\bf AVC-DNN, $f_{\text{DNN-3}}(\cdot)$}, which receives the LV side bus voltage information over $(k-1)T_c$ to $kT_c-T_s$ and outputs the predicted LV side voltage from $kT_c$ to $(k+2)T_c-T_s$ without assuming the influence of any future control action at $k+1$. 
With regards to the LTC control action at $k$, it is the same as the one decided at $k-2$, which is added to the stored LTC control for nominal model at $k$, yielding $\Delta u^{\text{LTC}}_k$. According to the above discussion, (\ref{pertmpcV2}) is expanded to show the contribution of the individual controls in (\ref{pertmpcV2mod}) below, where $\Delta u^{\text{LTC}}_k$ is known from beforehand (at $k-2$), whereas $\Delta u^{\text{SVC}}_k,\;\Delta u^{\text{LS}}_k$ are computed by a single step MPC optimization: 
\begin{multline}\label{pertmpcV2mod}
\hat V_{k:k+1} = \bar V^{\text{nom}}_{k:k+1} + \bar S^{\text{nom,\rm SVC}}_{k:k+1}\Delta u^{\text{SVC}}_k + \bar S^{\text{nom,\rm LS}}_{k:k+1}+\Delta u^{\text{LS}}_k \\ \;+\; \bar S^{\text{nom,LTC}}_{k:k+1}\Delta u^{\text{LTC}}_k
\end{multline}
The block diagram of this entire scheme of perturbation control computation is shown in Figure \ref{f3}.

\section{{Sensitivity Computations for Offline MPC}}\label{sectscomp}
This section presents the trajectory sensitivity computation part required to (i) solve the MPC problem offline for nominal system, and (ii) generate the training data set for sensitivity-DNN $f_{\text{DNN-2}}(\cdot)$. We used power system simulator PSAT \cite{milano2005open} for our studies, and found that the trajectory sensitivity computation with respect to control is not supported in it. There exist some earlier works \cite{jin2009model,guanji,guanji2} that mention the use of PSAT for trajectory sensitivity based MPC, but the details for the trajectory sensitivity computation are not provided. So one needs to essentially introduce certain extension to the source code of PSAT to achieve the trajectory sensitivity computation. Such PSAT extension required was first reported in our conference paper \cite{rrh2}, providing its detailed explanation; the summary presentation below is included for completeness.

As discussed earlier, sensitivity can be obtained solving (\ref{systs1})-(\ref{systs2}) for which we need the Jacobians $f_x$, $f_y$, $g_x$, $g_y$, $f_u$, $g_u$ at each time steps of the simulation. The issue is that PSAT does not support the computation of $f_u$ and $g_u$, and also does not store the values of the other Jacobians. For the latter, we introduced additional coding in PSAT's time domain integration subroutine to store the Jacobians for each step of time domain simulation. On the other hand, to overcome the lack of support for the computation of Jacobians $f_u$ and $g_u$, we utilized the fact that PSAT supports the computation of Jacobians with respect to state variables. Hence, we treated the control $u$ as a state variable, but having zero-dynamics $(\dot u =0)$ (recall that for computing sensitivity with respect to control, the control is held constant at its nominal value, i.e., is rate of change is indeed zero). With this augmentation of states to include zero-dynamics input, we computed the Jacobians $f_u$ and $g_u$ as follows. Denoting the control-augmented state variables as, \(\overline {x}= \begin{bmatrix} x\\
u \end{bmatrix}, \) algebraic variables as $y$, and combining, we have:
\begin{equation}\label{sysdaeaug}
\dot{\overline{x}} =  \begin{bmatrix}
\dot{x} \\
\dot{u} 
\end{bmatrix}
=
\begin{bmatrix}
f(x,y,u) \\
0
\end{bmatrix} =: \overline {f}(\overline{x},y);\quad
 0=g(\bar x, y)
       \end{equation}
\noindent Upon differentiation of equations (\ref{sysdaeaug}) with respect to control input $u$, we get:
\begin{equation}\label{systsaug}
  \dot{\overline{x}_u}(t)=\overline{f}_{\overline{x}}{\overline{x}_u}(t)+\overline{f}_y{y_u(t)};\quad
  0=g_{\overline{x}}{\overline{x}_u}(t)+g_y{y_u(t)}.
       \end{equation}
Note $\overline{f}_{\overline{x}}$ = $\begin{bmatrix}
f_x & f_u \\
0 & 0 \end{bmatrix}$, $\overline{f}_y$ = $ \begin{bmatrix}
f_y\\
0 \end{bmatrix}$, and $g_{\overline{x}}$ = $ \begin{bmatrix}
g_x & g_u
\end{bmatrix}$. Now with the above control-augmented states, PSAT can compute as well as store the Jacobians $\bar{f}_{\bar{x}}$, ${\bar{f}_y}$, $g_{\bar{x}}$, and $g_y$ in course of the time domain simulation, from which we extract the desired $f_u$ and $g_u$. Having all the Jacobians $f_{x}, f_y, g_{x}, g_y, f_u, g_u$ available, equations (\ref{systsaug}) are solved numerically to obtain the required trajectory sensitivities $x_u$ and $y_u$. Further details can be found in \cite{rrh2}. 

The default PSAT dynamic models of SVC and LTC are given by (\ref{svc}) and (\ref{ltc}), respectively:
\begin{subequations}
\begin{align}
 \dot{b}_{SVC}=\frac{K_r(V_{ref} - V) -b_{SVC}}{T_r}, \label{svc} \\
 \dot{m}=-K_dm + K_i(V - V_{ref}), \label{ltc}
       \end{align}
\end{subequations}
where the susceptance $b_{SVC}$ and tap-ratio $m$ are the control inputs, respectively. To make these models suitable for sensitivity computation, we chose the certain parameters of existing SVC and LTC blocks appropriately: In SVC, to zero the ${b}_{SVC}$ dynamics, we set he time-constant $T_r$ to be very high, whereas set the gain $K_r$ very low. Similarly, by assigning very low values to parameters $K_d$ and $K_i$, we also set the dynamics of $m$ in LTC block to close to zero. Certain other adjustments in the PSAT code were performed to account for the fact that these blocks also have anti-windup limiters.

The computation of load-shedding sensitivity is little more involved. In this study, we used exponential recovery load for which the active power dynamics is given by:
\begin{subequations}
\begin{gather}
 \dot{x}_{P}= -x_P/T_P + P_0(V/V_0)^{\alpha_s} - P_0(V/V_0)^{\alpha_t}, \label{ls} \\
P = x_P/T_P + P_0(V/V_0)^{\alpha_t}.
       \end{gather}
\end{subequations}
\noindent Here $P_0$ is the base load value that needs to be altered to exercise load-shedding. So we introduced an additional equation $\dot{P_0}=0$ to augment $P_0$ as another zero dynamics state-variable. Similar state-variable augmentation was done for base reactive power $Q_0$. To introduce the new state variables, we made certain modifications in the corresponding sub-routine of PSAT for the exponential recovery load. 

\section{Implementation, Test Cases, and Results}

As a proof-of-validation, our proposed methodology is implemented in PSAT and applied for voltage stabilization in IEEE-9 and IEEE-39 bus systems. We modified both these test systems to accommodate different control inputs and distribution-side loads through LTC.
\subsection{Test System 1: IEEE 9-Bus System}
We utilized the standard IEEE 9-bus system with slight modification (see IEEE 9-bus example in Figure \ref{fsys})  to address the voltage stability problem following a 3-phase fault at bus-5 with a fault lasting 0.10 sec, that got cleared by tripping the line between bus 4 and 5. In order to include the distribution-side loads, we added 2 additional buses 10 and 11, connected through LTC to bus 6 and 8, respectively, or the original 9-bus system. These 2 LTCs are controlled by their local AVCs. The other control inputs include 3 SVCs connected at buses 5, 7 and 8, each varying from 0 to 0.2 p.u. per step, and load-shedding of up to 0.1 p.u. (approx. 10\% of the bus load) at buses 10 and 11. A sustained under-voltage conditions are observed following the fault without any control actions, and eventually this leads to system collapse after 70 sec. of the fault occurrence. 

\subsection{Test System 2: IEEE 39-Bus System}
\begin{figure}[htbp]
  \centering
    \includegraphics[scale=0.60]{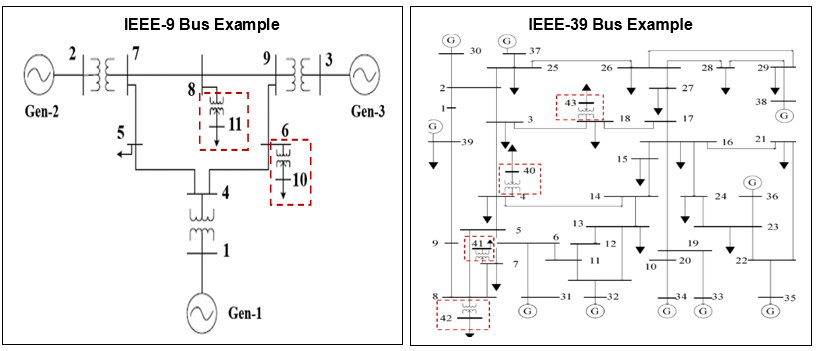}
  \caption{IEEE 9-bus and 39-bus systems.}
  \label{fsys}
   \vspace*{-0.2in}
 \end{figure}

For the IEEE-39 bus system (see IEEE 39-bus example in Figure \ref{fsys}), we consider the 3-phase fault at bus-15, which is cleared by tripping the transmission line in between bus 15 and 16 within 0.10 sec of the fault occurrence. Here, the added 4 buses, 40, 41, 42 and 43, are used to represent the loads of the distribution side, that are connected through 4 LTCs to the buses 4, 7, 8 and 18, respectively of the original network. In this power system, the SVCs are located at the buses 4, 5, 7, 8, 15, 17, 18 and 25 with value ranging from 0 to 0.20 p.u. per step, whereas the load-shedding can be exercised at the buses 15, 40, 41, 42 and 43 by up to 0.3 p.u. per step (approx. 10\% of the bus load). Like the IEEE 9-bus example, following the fault, voltage drops below the desirable level almost immediately, and hence, it needs to be stabilized to avoid any potential collapse.
\comment{\begin{figure}
     \centering
     \begin{subfigure}[b]{0.25\textwidth}
         \centering
         \includegraphics[width=\textwidth]{ieee-9.png}
         \caption{IEEE-9 Bus Example}
         \label{fig:9bus}
     \end{subfigure}
     \hfill
     \begin{subfigure}[b]{0.25\textwidth}
         \centering
         \includegraphics[width=\textwidth]{ieee-39.png}
         \caption{IEEE-39 Bus Example}
         \label{fig:39bus}
     \end{subfigure}
\end{figure}}
\subsection{Offline MPC Computed Controls for Nominal Models}
\begin{figure}[htbp]
  \centering
    \includegraphics[scale=0.48]{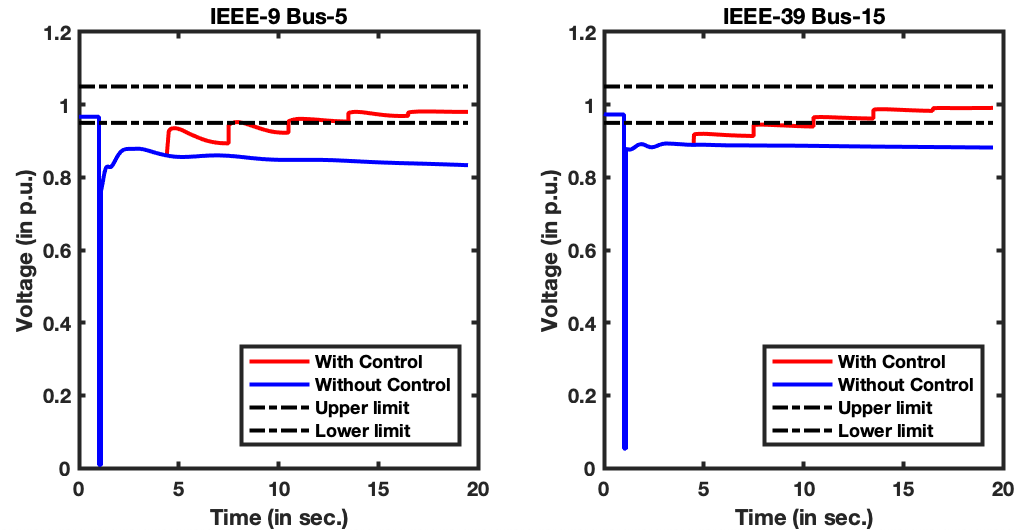}
  \caption{Voltage profile with ({\bf red}) and without ({\bf blue}) MPC for nominal model of IEEE 9-bus and 39-bus systems.}
  \label{voloff}
   \vspace*{-0.2in}
 \end{figure}

We utilized nominal models of the above 2 systems to compute their MPC control sequence $u^*_{\text{nom,seq}}$ offline and stored those values. In this computation, sampling interval $T_s=0.1$ s, whereas the control intervals $T_c$ are 3~s. The computed optimal control sequences for the respective nominal models are listed in Tables \ref{tab:result9} and \ref{tab:result39}, which are corrected online during runtime operation. The voltage trajectories with MPC control ({\bf red}) and without MPC control ({\bf blue}) MPC for bus-5 of 9-bus system and bus-15 of 39-bus system are shown in Figure \ref{voloff}, which clearly show the need of voltage stabilizing control in order to maintain voltages within the safe limits, usually [0.95,1.05] p.u.
\begin{table}[ht]
\caption{IEEE 9-Bus: Optimal sequence for nominal model}\label{tab:result9} 
\vspace{-0.5mm}
\centering
\tabcolsep=0.05 cm
\begin{tabular}{|c|c|c|c|c|c|c|c|}  
\hline 
Time instant  & 4.5 sec & 7.5 sec. & 10.5 sec. & 13.5 sec. & 16.5 sec.\\ [0.5ex] 
\hline 
SVC-5 (in p.u.) & 0.2 & 0.2 & 0.2 & 0.1378 &  0.0534 \\ 
\hline
SVC-7 (in p.u.) & 0.2 & 0.1791 & 0.0 & 0.0 &  0.0 \\
\hline
SVC-8 (in p.u.) & 0.2 & 0.0 &  0.0 & 0.0 &  0.0 \\
\hline
L/S 10 (in p.u.)& 0.0619 & 0.0024 & 0.0167	& 0.0 &  0.0 \\
\hline
L/S 11 (in p.u.)& 0.0353 & 0.0 & 0.0	& 0.0 &  0.0 \\
\hline
LTC b/w 10-6 & 0 & 0 & 0	& 0 &  0 \\
\hline
LTC b/w 11-8 & 0 & 0 & $+1$	& 0 &  0 \\
\hline
\end{tabular}
\vspace*{-.15in}
\end{table}

\begin{table}[ht]
\centering
\caption{IEEE 39-Bus: Optimal sequence for nominal model}\label{tab:result39} 
\vspace{-0.5mm}
\tabcolsep=0.05 cm
\begin{tabular}{|c|c|c|c|c|c|c|c|}  
\hline 
Time instant  & 4.5 sec & 7.5 sec. & 10.5 sec. & 13.5 sec. & 16.5 sec.\\ [0.5ex] 
\hline 
SVC-4 (in p.u.)& 0.2 & 0.2 & 0.2 & 0.2 &  0.0 \\ 
\hline
SVC-5 (in p.u.)& 0.2 & 0.2 & 0.2 & 0.2 &  0.0 \\
\hline
SVC-7 (in p.u.)& 0.2 & 0.2 &  0.2 & 0.2 &  0.134 \\
\hline
SVC-8 (in p.u.)& 0.2 & 0.2 & 0.2 & 0.2 &  0.2 \\
\hline
SVC-15 (in p.u.)& 0.2 & 0.2 & 0.2	& 0.2 &  0.2 \\
\hline
SVC-17 (in p.u.)& 0.0 & 0.0 & 0.0	& 0.0 &  0.0 \\
\hline
SVC-18 (in p.u.)& 0.2 & 0.2 & 0.0	& 0.0 &  0.0 \\
\hline
SVC-25 (in p.u.)& 0.0 & 0.0 & 0.0	& 0.0 &  0.0 \\
\hline
L/S 15 (in p.u.)& 0.3 & 0.3 & 0.3	& 0.3 &  0 \\
\hline
L/S 40 (in p.u.)& 0.3 & 0.3 & 0.3	& 0.3 &  0 \\
\hline
L/S 41 (in p.u.)& 0.3 & 0.3 & 0.3	& 0.3 &  0 \\
\hline
L/S 42 (in p.u.)& 0.3 & 0.3 & 0	& 0 &  0 \\
\hline
L/S 43 (in p.u.)& 0.3 & 0.3 & 0.3	& 0 &  0\\
\hline
LTC b/w 40-4 & 0 & 0 & $+1$	& 0 &  $+1$ \\
\hline
LTC b/w 41-7 &  0 & 0 & $+1$	& 0 &  0 \\
\hline
LTC b/w 42-8 &  0 & 0 & $+1$	& 0 &  $+1$\\
\hline
LTC b/w 43-18 &  0 & 0 & $+1$	& 0 &  0 \\
\hline
\end{tabular}
\vspace*{-.15in}
\end{table}

\subsection{DNNs for Online MPC-based Adaptive Control Correction}
The real-time application of our methodology relies on 3 categories of DNNs as mentioned in Section \ref{secpertcontrol}, namely Prediction-DNN, Sensitivity-DNN, and AVC-DNN. The training of these DNNs is an important factor for the success of the proposed methodology. For both IEEE 9-bus and 39-bus cases, we created a large pool of training data by simulating the respective systems under the influence of the offline computed optimal control sequence for the nominal models, while changing the initial loads to within $\pm 20\%$ of nominal initial load. This data ensures the exploration of the search space is around the nominally controlled model trajectory. For both systems, we randomly created 2500 different load levels, and collected approx. 12000 and 15000 unique data samples for training of Prediction-DNN and Sensitivity-DNN, respectively. 
Similarly to train the AVC-DNN,  
we considered 1500 different initial load conditions and gathered 6000 training data. 
We divided all these three categories of data into $70:30$ ratio to create the training versus the test data sets for the three DNNs. 

We used a feed forward structure with 2 hidden layers to build the three DNNs. The number of neurons of the hidden layers were chosen to be between 64 and 256, and $\tanh(\cdot)$ was used as the activation function. 
The optimizer chosen for the training is ADAM, with gradient momentum $\beta_1=0.9/0.95$ and RMS momentum $\beta_2=0.999/0.95$. The loss function, batch size and learning rate are: mean squared error (MSE) loss, 32, and $10^{-3}$, respectively. Standard techniques to avoid over-fitting and facilitate fast learning were practiced: (i) adding drop-out layers, and (ii) normalizing the inputs and outputs of the DNNs. 

 \begin{figure}[htbp]
  \centering
    \includegraphics[scale=0.50]{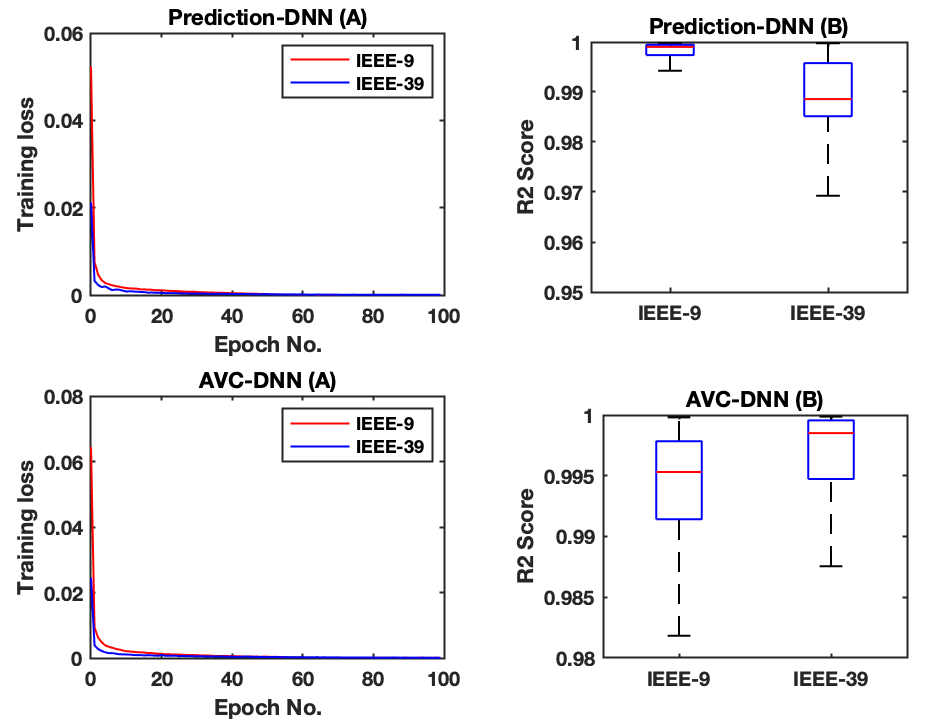}
  \caption{Performance of Prediction-DNNs and AVC-DNNs.}
  \label{f4}
 \end{figure}
\begin{figure}[htbp]
  \centering
    \includegraphics[scale=0.50]{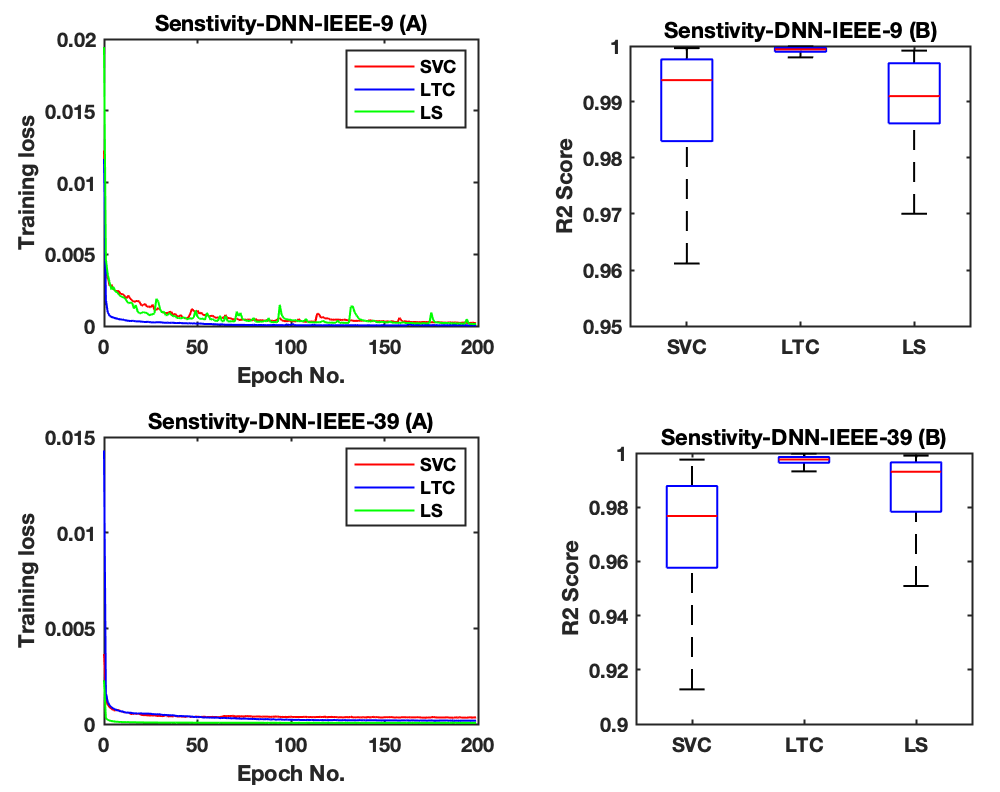}
  \caption{Performance of Sensitivity-DNNs.}
  \label{f5}
 \end{figure}

In Figures~\ref{f4}~and~\ref{f5}, the training performance is shown in terms of MSE, and the test performance is determined by measuring the coefficient of determination or $R^2\in[0,1]$ score of the DNN predicted value and the respective actual values over the test data sets. (A $R^2$ value of 1 indicates an exact fit.)
Figures~\ref{f4}~and~\ref{f5} confirm prediction accuracy of more than 95\% of the trained models, establishing that they offer good fit for the online adaptive control scheme proposed in this article.
\subsection{Online MPC-based Adaptive Control and its Robustness}
\begin{figure}[htbp]
  \centering
    \includegraphics[scale=0.50]{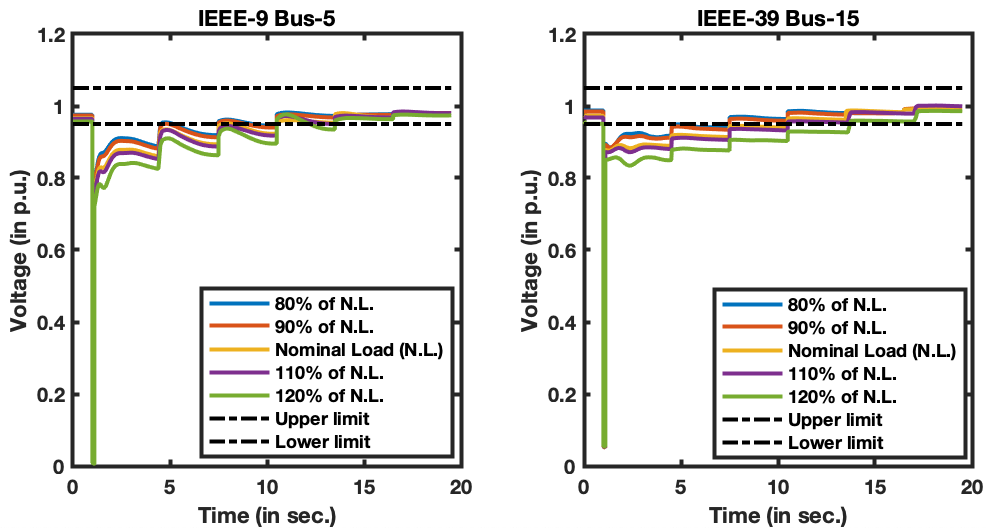}
  \caption{Voltage profile with online MPC-based adaptive control for IEEE 9-bus and 39-bus systems.}
  \label{fvolfinal}
    \vspace*{-0.1in}
 \end{figure}

\begin{figure}[htbp]
  \centering
    \includegraphics[scale=0.48]{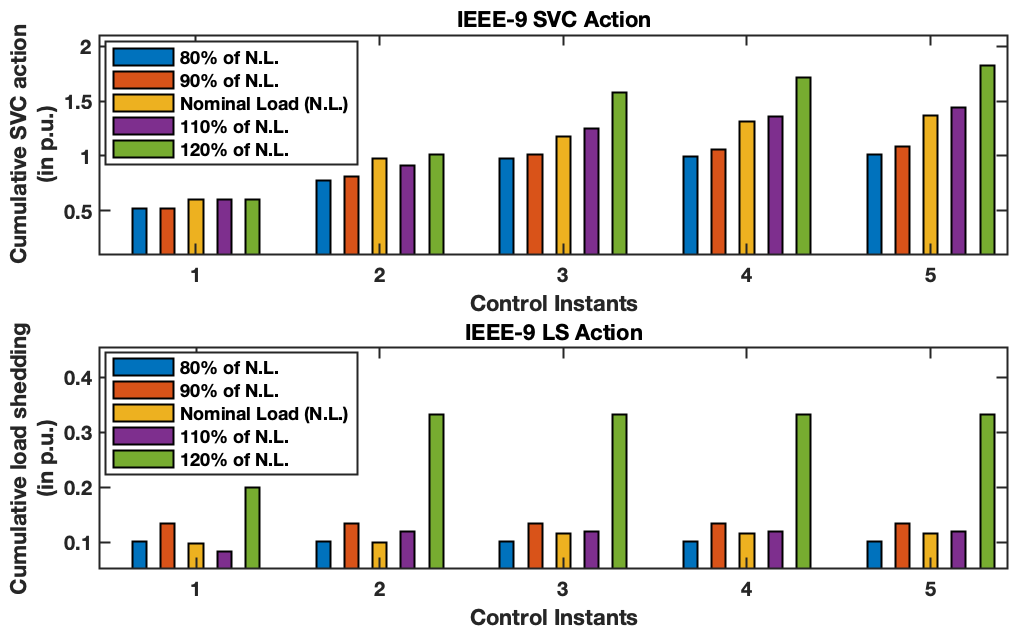}
  \caption{Computed SVC and LS controls for IEEE 9-bus system for 5 loading conditions.}
  \label{f9buscon}
 \end{figure}

\begin{figure}[htbp]
  \centering
    \includegraphics[scale=0.48]{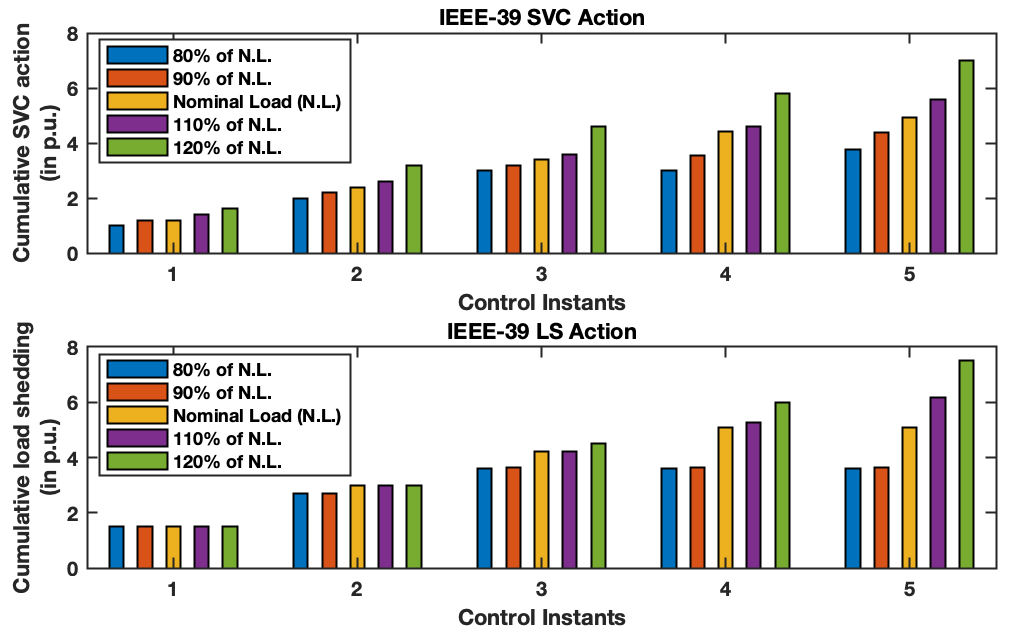}
  \caption{Computed SVC and LS controls for IEEE 39-bus system for 5 loading conditions.}
  \label{f39buscon}
 \end{figure}

The real-time performance of the proposed scheme is evaluated for both IEEE 9-bus and 39-bus systems under different load conditions. In each case, the real-time control corrections are computed based on current measurements and the offline computed respective optimal control sequences for the nominal load models. Due to space constraints, we describe 4 different load levels, 80\%, 90\%, 110\%, 120\% of the nominal load for showing the performance and robustness of the methodology. The voltage profiles for each of the above cases are shown in Figure \ref{fvolfinal}, indicating clearly that the proposed scheme is successful in restoring the desired voltage levels under different operating conditions (and so effectiveness and robustness of the proposed approach). 
To check the control inputs, we computed the total SVC and LS actions at each control instants, and plotted the cumulative sum of the respective actions in Figure \ref{f9buscon} and Figure \ref{f39buscon}. The trend suggests that with increase of load, the amount of controls introduced increased, which is as expected. 
 
Finally, and importantly, the average online computation times of the original MPC and the proposed online scheme are compared in Table \ref{tab:time}, which demonstrates the proposed scheme is $\sim$20-fold faster than the original offline computed MPC implementation, and takes under 0.3~s to compute a control at each online decision instant, which is under one-tenth the control interval, making MPC real-time and practical for power systems. It is important to contrast that even the traditional controllers, e.g., UVLS relaying scheme generally needs $\sim$0.5~s to decide a control action \cite{7822924}. For our implementation and computation, we used intel(R) Core(TM) i7-4790 CPU @ 3.60GHz processor with 16 GB RAM.  
\begin{table}[htbp]
\caption{Comparison of computation time}\label{tab:time} 
\vspace*{-.5mm}
\centering
\tabcolsep=0.5 cm
\begin{tabular}{| c | c | c |}
\hline
\textbf{Method} & \multicolumn{2}{ c| }{\textbf{Average CPU Time}}  \\ 
\cline{2-3}
& IEEE 9-bus & IEEE 39-bus \\
\hline
Original MPC & 4.50 sec/step & 7.00 sec/step  \\ \hline
Proposed Online Version & 0.27 sec/step & 0.29 sec/step\\ \hline
\end{tabular}
\vspace*{-.15in}
\end{table}

\section{Conclusions}\label{sec:conclusion}
The paper proposed a framework for {\em real-time} implementation of MPC in power systems for a first time. A combination of offline MPC-based control optimization for {\em nominal system}, and an iterative online control correction based on measurements of {\em real system} is proposed, where the online step is further sped through the introduction of trained DNNs for voltage trajectory prediction and its sensitivity estimation. By exploring the space in neighborhood of the nominal trajectory of offline computed control, the search space for DNN training was drastically reduced to make it practical. The test results applied to IEEE 9-bus and 39-bus systems show the remarkable performance of newly proposed scheme in terms of efficacy, robustness with respect to load variations, and online computation time, that has been reduced to less one-tenth the control interval (and comparable to traditional control computations), making the real-time implementation of the MPC practical. Future research directions can include quantification of resilience indices \cite{soumyo} of the MPC-controlled system that may also involve reachability study \cite{jin2010reachability}.


%



%

\bibliographystyle{IEEEtran}
\bibliography{IEEEabrv,Bibliography}

%




\end{document}